\documentclass{article}
\usepackage{url}
\usepackage{moreverb}
\usepackage{graphicx}
\usepackage{color}
\usepackage{multirow}
\usepackage[left=3.5cm,top=3cm,bottom=2cm,right=2.5cm]{geometry}

\begin{document}

\title{When Should I Use Network Emulation?}

\author{Emmanuel Lochin$^{1,2}$, Tanguy P\'{e}rennou$^{1,2}$, Laurent Dairaine$^{2}$\\
~\\
$^1$ CNRS ; LAAS ; 7 avenue du colonel Roche, F-31077 Toulouse, France \\
$^2$ Universit\'{e} de Toulouse ; UPS, INSA, INP, ISAE ; LAAS ; F-31077 Toulouse, France \\
\texttt{firstname.lastname@isae.fr}
}

\date{}

\maketitle

\begin{abstract}
The design and development of a complex system requires an adequate methodology and efficient instrumental support in order to early detect and correct anomalies in the functional and non-functional properties of the tested protocols. Among the various tools used to provide experimental support for such developments, network emulation relies on real-time production of impairments on real traffic according to a communication model, either realistically or not. 

This paper aims at simply presenting to newcomers in network emulation (students, engineers, ...) basic principles and practices illustrated with a few commonly used tools. The motivation behind is to fill a gap in terms of introductory and pragmatic papers in this domain. 

The study particularly considers centralized approaches, allowing cheap and easy implementation in the context of research labs or industrial developments. In addition, an architectural model for emulation systems is proposed, defining three complementary levels, namely hardware, impairment and model levels.  With the help of this architectural framework, various existing tools are situated and described. Various approaches for modeling the emulation actions are studied, such as impairment-based scenarios and virtual architectures, real-time discrete simulation and trace-based systems. 
Those modeling approaches are described and compared in terms of services and we study their ability to respond to various designer needs to assess when emulation is needed.
\end{abstract}


\section{Introduction}
\label{sec:intro}

Designing and developing communication protocols and real-time systems is a complex process where various actors participate in different phases, having only a partial vision of the whole system. The experiment phase, which eventually provides a global vision, is a mandatory step in research and development process of distributed applications and communication protocols. In this context, three classical ways to achieve experimentation are commonly used: simulation, live testing and more recently emulation.

Simulation, particularly event-driven simulation, is a classical way to achieve economical and fast protocol experimentation. It relies on an ad-hoc model to work with and it uses a logical event-driven technique to run the experiment.  The use of modeling techniques simplifies the studied problem, by concentrating on the most critical issues. Indeed, network simulators are essential to provide a proof of concept prior to protocol development.
Nevertheless, those tools (based on a virtual clock) cannot replace practical protocol evaluation that quantifies implementations' overhead during real-time operation.
Eventually, to realize a real-time evaluation, only two solutions are left: live testing and network emulation. 
For instance, network simulation has been used to design the TFRC protocol internal mechanisms \cite{rfc3448} (the TCP-Friendly Rate Control protocol is a rate-based congestion control mechanism for unicast flows operating in a best-effort Internet environment); then, a user-level prototype has been realized \cite{tfrc-code} to quantify the processing overhead related to the inherent implementation. Throughout the remainder of this paper, we use the TFRC case as a running example to help the understanding of certain concepts presented. 

In live testing, evaluations are driven with real implementations. The fundamental way to do the experiment is by using real technology for the underlying networking environment. This real environment can be the target network or an ad-hoc testbed involving real equipments. Nevertheless, this approach is considered to be very expensive and inflexible to evaluate all aspects of the protocol being tested.

Emulation is considered to be at the cross-road between simulation and live testing. This approach consists in executing and measuring real protocols and application implementations over a certain network where part of the communication architecture is simulated in real-time. The aim of emulation is to allow a distributed software to run either in realistic conditions (e.g. over a satellite network) or specific conditions (e.g. when specific packets are dropped such as SYN packets in the TCP case).

This paper introduces an overview of existing network emulation approaches. We particularly focus on the study of centralized approaches, allowing simple implementation in the context of research labs or industrial development centers. In addition, we propose a general architectural model that allows various emulation approaches to be presented and situated in the model. These emulation approaches include mainly impairment scenarios and virtual architectures. Furthermore, a comparison of these approaches and a set of criteria considered as requirements for emulation systems will be proposed.

This paper is organized as follows: section \ref{NetworkExperimentation} situates emulation among various experimentation approaches. Section \ref{EmulationRequirements} presents main emulation requirements. Additionally, section \ref{NetworkEmulationModel} discusses network emulation architecture based on three complementary levels. The outline of main emulation approaches, as well as their positions in the architectural model proposed will be discussed in section \ref{EmulationApproaches}.  Finally, some concluding remarks are given in section \ref{Conclusion}.

\section{Network Experimentation \\Approaches}
\label{NetworkExperimentation}

Before diving in the world of emulation, we first present the common experimental approaches used in research labs and industrial development.

\subsection{Simulation}
As discussed above, simulation is a very effective and efficient way to experiment with protocols. Network simulation typically utilizes ad-hoc model and logical event-driven techniques. Classical tools such as ns-2 \cite{ns-2} or OPNET \cite{OPNET.2001.OPNET} provides a core simulation engine, as well as a large set of protocol models. These simulation tools allow experiments to be done without high costs involvement. The modeling techniques used in the simulators allow the studied problems to be simplified by concentrating on the most important issues. Furthermore, simulation tools do not operate in real-time. Therefore, depending on the model complexity, it is possible to either simulate a logical hour in few real-time milliseconds or a logical second in several real-time days. This characteristic reflects both the benefit and weakness of simulation tools. Due to this attribute, it is unfeasible for simulation tools to implement systems involving man-in-the-loop. Most simulation tools do not allow to test real-time implementations but only models (even the most innovative and sophisticated one).

\subsection{Simulation shortcomings}

As already emphasized in the introduction, network simulators are essential to provide a proof of concept prior to protocol development but cannot replace practical protocol evaluation. To pursue with the TFRC example previously introduced in Section \ref{sec:intro}, the concept of the rate-based equation has been validated within ns-2 simulator while the feasibility of the implementation has been evaluated through real-time experiments. Even though the core algorithm developed within ns-2 in C++ has been reused inside the kernel implementation, most of the data structure, message exchange and protocol framework had to be written from scratch. This additional code has to be evaluated too.

Furthermore, it is important to ensure that the services and performances offered by the simulation model are consistent with the real experimental implementation of the protocol.

\subsection{Live experimentation}

Another conventional method to test and debug distributed software during the implementation stage is to use real hardware and/or software components. The software can be tested either on a real target network or on an ad-hoc testbed using real equipment. However, this approach is particularly expensive in the context of wide area networks, especially when using specific technology such as satellite network. The cost inefficiency of this method does not involve only the technology cost but also the distributed man-in-the-loop manipulations and synchronization required. Moreover, it is sometimes impossible to use this approach simply because the new technology support is not yet validated or available, e.g. when developing an application over a new satellite transmission technology that is not yet operating. This method also suffers from the inherent discrepancies between a particular test network and the much broader range of network imperfections that will be encountered by the software users. 

Using real technology on target operational network has been widely deployed. An example of this scheme is well illustrated by PlanetLab \cite{Peterson02Blueprint}. PlanetLab is a distributed platform that alleviates experiments management, offering a way to use a very large set of hosts over the Internet. However, the purpose of PlanetLab is to use Internet as a testbed and not to control the network experimentation conditions. As a result, PlanetLab does not target reproducibility and is thus more efficiently used for metrology experiments. 

\subsection{Network Emulation}

Since several years, progresses in high speed processing and networking have allowed the rapid development of network emulators, such as Dummynet~\cite{Rizzo.1997.Dummynet}, NIST Net~\cite{Carson.2003.NIST}. Network Emulation is a weighted combination of real technology and simulation. It is used to achieve experiments using both real protocol implementations and network models. Basically, this allows the creation of a controlled communication environment.  This communication environment can produce specific target behaviors in terms of quality of service. The objective of emulation tools is to reproduce a real underlying network behavior, such as configurable wired \cite{Zec.2004.Operating} or wireless \cite{Zheng.EMPOWER.2003}, \cite{wnine} topologies. Additionally, emulation aims at providing ``artificial impairments'' on the network to test particularities of the experimented protocol. These impairments include loosing specific packets, reducing the network bandwidth with a specific timing or introducing delay over the network. Emulation is particularly useful in the debugging and testing phase of a system. 

\section{When do I need emulation?}
\label{EmulationRequirements}

You need emulation to assess the performance of an end-to-end system. Although you can use emulation at any layer of the OSI model, in the present paper, we focus on network emulation which is a combination of real technology (application and communication stack above link-level) and simulation of the behavior of the link and physical levels.

Let's assume you want to assess the performance of the TFRC transport protocol. You might be interested in validating the use of your implementation over several types of terminal (\emph{e.g.}  mobile phone, PDA, laptop, server, etc.) and compare whether the impact of TFRC internal algorithms behave similarly in various network conditions. In order to drive this test, you must evaluate as a non-exhaustive list: the memory footprint, CPU usage and packet processing overhead to identify potential limits and propose implementation and algorithmic improvements. These metrics are not available in a simulation context. Testing TFRC with various bandwidth size in a real setup would involve the use of several different network setups (i.e. several testbeds with different cards on different hosts) while emulation provides an easy way to set the bandwidth of a link. Anyway, network emulation is the most practical scheme to obtain trustable metrics since you change only one parameter of your experimental setup.

In order to assess Quality of Service (QoS), the overall performances obtained from an end-to-end protocol are mainly dependent on external factors such as underlying technologies (\emph{e.g.} RSVP establishment path, Service Level Agreement with a DiffServ network, etc.), interconnection topologies, current network traffic and so on. Different types of QoS can also be offered by the underlying network. 
For example, IP network service could offer a communication channel ranging from perfect (minimum delay, high bandwidth and no packet loss, as in a gigabit LAN) to unsatisfactory (high delay, low bandwidth and high packet loss rate (PLR), as in a noisy satellite network), depending on the underlying protocols and many other external factors. This leads to a large set of possibilities in the protocol experiments that can be created. The different types of end-to-end QoS that can be produced by the underlying experiment framework can focus on:

\begin{itemize}
    \item   \emph{Artificial QoS}: the experiment framework provides a way to evaluate the protocol over specific QoS conditions, not imperatively related to any technology or realistic conditions. Artificial QoS allows the user to test and focus on its experimental protocol in target QoS conditions. This can be considered as a form of unit testing. Furthermore, the aim of this method is to point out errors or bugs that are difficult to observe in a non-controlled environment where they rarely happen. This can be used, for instance, at the transport level to study the impact of various packet drops in a TCP connection (\emph{e.g.} SYN/ACK packets~\cite{pdcn06}, etc.). 
    \item   \emph{Realistic QoS}:  the experiment framework provides a way to reproduce the behavior of some specific network architecture as accurately as possible. This type of experiment allows the user to evaluate the protocol over an existing network or inter-network without using a real testbed and all related technologies (e.g. a wireless network, a satellite network, an Ethernet gigabit network, or any interconnection of such technologies).
\end{itemize}

Generally the following set of impairments are commonly at least supported by almost all emulators systems: round trip time delay, jitter, packet loss rate and bandwidth size.

Today, there are several emulation platforms freely available on the Internet, either remotely accessible (\emph{e.g.} EmuLab~\cite{White.2002.Emulab}, Orbit~\cite{orbit}, \cite{orbit2}) or for download and local installation (\emph{e.g.} Imunes~\cite{Imunes}, Netem~\cite{netem}, Dummynet~\cite{Rizzo.1997.Dummynet}, KauNet~\cite{mobieval2007}).  
We strongly believe it would be not appropriate to simply list and detail all these proposals. Instead, we propose in the following Section an architectural model where essential features are highlighted.

\section{Network Emulation \\Architectural Model}
\label{NetworkEmulationModel}

\begin{figure}[htbp]
\begin{center}
\includegraphics[height=8cm]{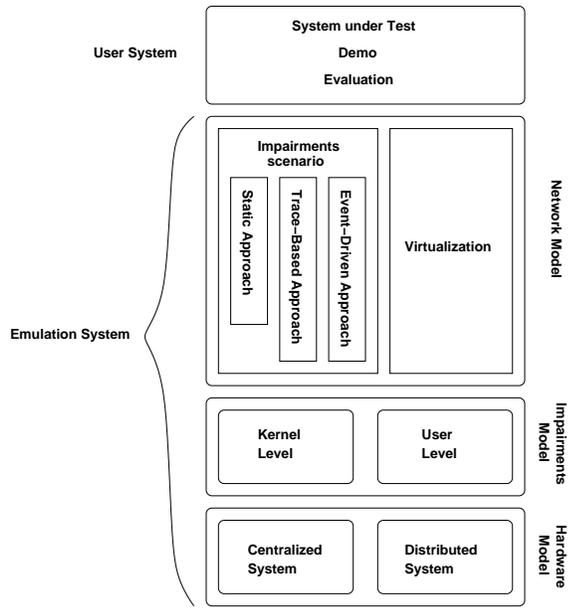}
\caption{Architectural model for emulation systems.}
\label{fig:ModelEmulation}
\end{center}
\end{figure}

Network emulation systems are based on various conceptual levels as illustrated in Figure \ref{fig:ModelEmulation}.
In this figure, we split an emulation system into three complementary levels, denoted Model Level, Impairment Level and Hardware Level. Each of these levels will be discussed in more details. Note that the User System is not considered to be a part of the emulation system. It includes the System Under Test, for instance a protocol or a distributed application to be evaluated or demonstrated as well as traffic sources and sinks.

\subsection{Hardware level}

The lowest layer of the proposed architecture, namely \emph{hardware layer}, represents the physical devices really used by the emulation system. 
These devices comprise the real end-systems, the real network links that interconnect them and possibly, network components such as switches or routers.
The virtual resources of the rest of the Emulation System and the User System are mapped on those real resources, \emph{e.g.} several virtual end-systems can reside on a single real computer.
It is crucial to understand that hardware level is not necessarily composed by the technologies associated to the emulated network conditions. 
For instance, emulating a satellite link to evaluate the performance of the TFRC protocol can be roughly done over a few desktop stations interconnected with ethernet links by setting appropriate PLR and delay on the resulting emulated link (see Section~\ref{subsec:static}).

The emulation system itself can be based on either a centralized system or a distributed system. 
In a centralized emulation system, we only use one computer to host the sender(s), the receiver(s), the intermediate node(s) and to manage all the impairments which define the experiment; 
while a distributed emulation system uses several computers to realize the same task.
As an example, the Imunes~\cite{Imunes} system falls in the first category while Dummynet~\cite{Rizzo.1997.Dummynet} or EmuLab~\cite{White.2002.Emulab} belong to the second one.

The main advantage of using distributed rather than centralized emulation system is the computation efficiency. However, this can be also considered as a disadvantage as it requires more physical 
resources and then, is much more complex to manage and administrate.
For example, an EmuLab-like testbed requires at least five computers: a sender, a receiver, a core emulator and two computers used to emulate both links in order to drive an experiment. 
This raises the problem of time synchronization of all machines that can be solved by using the NTP protocol \cite{rfc1305}. Despite the use of NTP and in the context of delay estimations, computers can experience clock drift that might compromise the measurements.  
Although a distributed emulation system is greedy in terms of resources, it remains more appropriate to estimate overall resource usage consumed by the protocol itself as it isolates the protocol 
under test from the emulation system. The advantage of a centralized emulation system is that it shares the same clock for all its components and is inherently synchronized.

\subsection{Impairment Level}

The \emph{impairment level} provides a mean to introduce impairments over the exchanged packet flows. The impairment system is a center piece of the whole emulation because the real target network conditions are 
driven by the impairment system. The accuracy of the emulation is deeply associated to the capacity of impairment systems to process the packets in time and without introducing any other impairment than 
those specified in the upper level. For instance, the impairment processing overhead might bias the packets processing time estimation.

An impairment can be introduced at either the kernel level or the user level. An example of emulator that introduces impairment at the kernel level is Dummynet \cite{Rizzo.1997.Dummynet}. Dummynet intercepts packets at the IP forwarding level by implementing a queue (named pipe by Dummynet API) able to introduce impairments on the enqueued packets. Dummynet is configured through the FreeBSD firewall API where each pipe is set up as a simple forwarding rule. Another similar tool implemented inside the GNU/Linux kernel is NIST-Net \cite{Carson.2003.NIST}. While Dummynet employs sophisticated queuing models for bandwidth modeling, NIST-Net includes delay models of much statistical sophistication. Indeed, NIST-Net is able to implement a varying delay scenario according to a given distribution while Dummynet uses a static delay because of the use of a queue. Both emulators cover complementary needs.

Finally, an example of user level impairment is ONE \cite{ONE}. ONE provides similar capabilities as Dummynet at the user level.
However, the clock timer resolution is a function of the kernel configuration and in general it ranges from $1ms$ to $10ms$.
Indeed, several system scheduler runs at a default $100 Hz$, meaning times based on normal system calls cannot be more precise than 10 milliseconds. 
As a result, a user level emulation cannot be as accurate as a kernel level one which gets a granularity close to the nanosecond. However, this approach is simple to install and adapted for many simple educational purposes.

\subsection{Network Model Level}

The \emph{model level} defines two ways to control the emulation behavior. 
A \emph{user impairment scenario} consists of an explicit list of impairment events while a \emph{virtual network architecture} generates implicit impairment events based on the virtual topology, equipments, link 
characteristics, communication and routing protocols. 
Both models will be discussed in more details in the next section. 

\section{Emulation Approaches}
 \label{EmulationApproaches}

\subsection{Impairment scenario models}

There are various types of scenarios. They can be classified as impairment scenario models as described throughout the rest of this section.

\subsubsection{Static Approach}
\label{subsec:static}

In a static approach every parameter remains constant throughout the experiment. Therefore, the static settings need to be configured before the experiment is conducted. It does not describe the real network very accurately since the behavior of real network changes all over time. However, it is sufficient to reproduce pragmatic cases of artificial quality of service (\emph{e.g.} bounded delay which characterizes specific network such as satellite link). The parameters that can be defined statically include delay, packet loss rate, bit error rate (BER), packet reordering, etc. This emulation model is usually useful to test all the possibilities of a product or to compare it to other already existing products. This is the basic behaviour of emulator such as Dummynet~\cite{Rizzo.1997.Dummynet} which is mostly used in this way.

\subsubsection{Event Driven Approach}

The key idea in event-driven~\cite{Dawson.1995.EventDriven} approaches is to apply impairements according to events. The most commonly used events are clock ticks (time-driven approach), but other events can be used, such as packet numbers, specific conditions observed on the traffic or purely random occurences. The event-driven approach is very useful to schematically represent a general behavior. The tester will be able to validate the product under several conditions 
and to compare it to other solutions. This approach has been used in various emulation tools. For example, Net Shaper \cite{Herrscher.2002.NetShaper} uses time oriented emulation. In Net Shaper, a daemon is executed and that 
daemon would wait for the new model to be applied to the emulation processor. The daemon is able to successfully receive and process up to 1000 messages per second. 

In the case of clock tick events, all impairements are triggered at user-defined times. Such approaches may be enforced by scripts which list all time events and associated actions. Time-driven models allow user to define the network and to make it evolve with time. As an example of such emulators we can notice IREEL~\cite{ireel} and WNINE~\cite{wnine}. Both emulators use an XML script containing update messages for a Dummynet static emulator used as an impairement engine.  As an example, the emulated network can be designed to behave differently during the day and during the night. 

Packet numbers in a flow can also be used as events, as in the KauNet network emulator~\cite{mobieval2007}. In that case, the ipfw tool of FreeBSD is used to select a flow, and data-driven patterns are used to define how the impairments change with packet numbers. For instance, a packet-loss pattern will use zeros for packet drops and ones for packet deliveries. KauNet also supports bit-error patterns, bandwidth change patterns, delay patterns and reordering patterns. Such patterns can also be used in a time-driven way, thus offering a more classical time-driven approach.

Other types of events have been proposed. Randomly generated events can be used to emulate random node failures. With an emulator able to read packet contents, a specific content (I image) or header value (DCCP handshake) can be detected and used as a triggering event. More generally, the metrology of the traffic can be used to detect specific conditions, such as the amount of flow reaching some level, to trigger specific impairments such as halving the bandwidth on the link.

Note that the most natural way to use the event-driven approach is the use of scripts associating impairment parameters with events, either explicitly like XML scripts IREEL or WNINE, or implicitely like patterns and scenarios in KauNet.

\subsubsection{Trace-based Approach}

This approach \cite{Noble.1997.Trace-based} is more realistic because the behavior of the network is obtained and will be reproduced exactly in the same way. 

First, a collection phase is usually done by using probes. These probes are used to record the dates of packets arriving or leaving a host. The results are transmitted to a controller that evaluate the delay and the mean loss rate to give the basic network model. This allows the user to get dynamic network profile. The limitation of the trace-based approach is that it cannot reproduce all conditions a network would experience. A single trace can only capture a snapshot of the varying performance along a particular path. Furthermore, the traces cannot fully reproduce the network behavior because it is non deterministic. The same situation in another time could have produced different parameters.  

The advantages of the trace-based approach is to use existing traces representing complex mobility movement to evaluate a prototype \cite{haggle}, \cite{crawdad}.

\subsection{Virtualization}

Network and system virtualization allow to easily manage multiple networks and systems, each of them customized to a specific purpose at the same time over the same shared infrastructure \cite{virtual01}. Nowadays, virtualization is perceived as the best candidate to support multiple router software candidate releases simultaneously as a long-run testing method before real deployment in the Internet (see for instance \cite{virtual02}). However, in this paper we are more interested in the second role of virtualization which is to run simultaneously multiple experiments in a shared experimental facility. 
 
The virtual architecture models are higher level models allowing the representation of a target network that is going to be emulated. It consists of two different aspects namely System Emulation and Real-time Discrete event simulation. This allows the design of an emulation model according to a real network topology where the experimented flow crosses a set of real or virtual nodes. Another way to achieve this is to use the Real-time Discrete Event Simulation through the establishment of a bridge between real packets and simulated event-driven environments as in the ns-2 emulation extension NSE~\cite{Fall.1999.Network}. 

	The global network behavior is produced by virtually reproducing the network topology and components. Two directions are taken depending on the way this virtualization is achieved. In the virtual systems, all nodes constituting the target network to be emulated are implemented either onto a single centralized system (several virtual nodes co-exist into the centralized system) or distributed onto various distinct systems (\emph{e.g.} a computing grid) usually connected together by high speed networks. Virtual links are used to connect these nodes together according to the topology of the targeted network. Real protocols such as IP or routing ones can also be implemented into the virtual node system. Of course, in this type of architecture, the classical strategy to produce realistic behavior is to introduce real traffic into the emulated network to produce congestion, delays, losses, etc.

	Imunes \cite{Zec.2004.Operating} is an example of centralized virtual node approach. It proposes a methodology for emulating computer networks by using a general purpose OS kernel partitioned into multiple lightweight virtual nodes. The virtual nodes can be connected via kernel level links to arbitrarily form complex network topologies. Furthermore, Imunes allows to emulate fully functional IP routers over each emulated virtual nodes. Imunes provides each virtual node with an independent network stack, thus enabling highly realistic and detailed emulation of network routers. It also enables user-level applications to run within the virtual nodes. At user level, Imunes proposes a very convenient interface allowing to easily define the emulated network, namely the virtual nodes, the software, the links and the impairment parameters. 

	The Entrapid protocol development environment \cite{Huang.1999.Entrapid} introduced a model of multiple virtualized networking kernels, which presents several variants of the standard BSD network stack in multiple instances, running as threads in specialized user process. Other approaches that following this approach is the Alpine emulator \cite{Savage.1999.ALPINE} project; GNS3/dynamips \cite{gns3} and Virtual Routers \cite{Baumgertner.2002.VirtualRouters}.

	The virtual architectural approach is often considered as the only mean to achieve realistic emulation of complex network topology. As previously introduced, PlanetLab proposes to directly use the Internet links to obtain 
real measures (for metrology purpose) conjointly with an emulation system allowing to map several end-hosts on a single computer in order to drive several and different experiments in parallel.
Nevertheless, the major problem of this approach may be the scalability issue. Issues such as how to implement one or several core network routers in a single machine and how to manage the number of flows in a centralized manner remain problematic. These questions are difficult to answer, not only in the context of total centralization but also in the context of distributed systems such as grids. 

\section{Summary}

We propose in this section a summary of the main characteristics of the emulators cited in this paper (following our classification model presented in Figure \ref{fig:ModelEmulation}). Note that the trace-based approach is a functionnality that is already included inside some emulators (Orbit, W-NINE, KauNet, ...) and can be added as a preprocessing tool to any event-driven emulator. However, we do not list this capability when it corresponds to an option and is not used as default network model.   

\begin{table}[ht!]
\begin{center}
\begin{tabular}{ l|c c c c c}\hline
\multirow{3}{2.5cm}{Name} 				&Hardware 	&Impairments		&Network 	&Year	& Web	\\
							&Model		&Model			&Model		&	&	\\ 
							& -- comments	&			&		&	&	\\ 
\hline
\hline
\multirow{2}{2.5cm}{Dummynet \cite{Rizzo.1997.Dummynet}}	&centralized	&kernel			&static		&1997	&\cite{dummynetweb}	\\
							&\multicolumn{5}{l}{-- FreeBSD, IP-level emulation} 						\\ \hline
\multirow{2}{2.5cm}{KauNet \cite{pdcn06}}		&centralized	&kernel			&event-driven	&2006	&\cite{kaunetweb}	\\
							&\multicolumn{5}{l}{-- based on dummynet} 							\\ \hline
\multirow{2}{2.5cm}{IREEL \cite{ireel}}			&distributed	&kernel			&event-driven	&2006	&\cite{ireelweb}			\\
							&\multicolumn{5}{l}{-- based on dummynet} 							\\ \hline
\multirow{2}{2.5cm}{Netem \cite{netem}}			&centralized	&kernel			&static		&2005	&\cite{netemweb}	\\
							&\multicolumn{5}{l}{-- GNU/Linux, IP-level emulation} 			\\ \hline
\multirow{2}{2.5cm}{NISTnet \cite{Carson.2003.NIST}}	&centralized	&kernel			&static		&2003	&\cite{nistnetweb}	\\
							&\multicolumn{5}{l}{-- sophisticated statistical distributions} 		\\ \hline
\multirow{2}{2.5cm}{ONE \cite{ONE}}			&centralized	&user			&static		&2001	&\cite{ONE}	\\
							&\multicolumn{5}{l}{-- the first network emulator} 			\\ \hline
\multirow{2}{2.5cm}{PlanetLab \cite{Peterson02Blueprint}}	&distributed	&kernel			&virtualization	&2003	&\cite{planetlabweb}	\\
							&\multicolumn{5}{l}{-- based on Linux virtual machines} 			\\ \hline
\multirow{2}{2.5cm}{EmuLab \cite{White.2002.Emulab}}	&distributed	&kernel			&virtualization	&2001	&\cite{emulabweb}	\\
							&\multicolumn{5}{l}{-- based on dummynet links} 				\\ \hline
\multirow{2}{2.5cm}{Imunes \cite{Imunes}}			&centralized	&kernel			&virtualization	&2003	&\cite{imunesweb}	\\
							&\multicolumn{5}{l}{-- based on FreeBSD virtual machines} 			\\ \hline
\multirow{2}{2.5cm}{Alpine \cite{Savage.1999.ALPINE}}	&centralized	&kernel			&virtualization	&2001	&No	\\
							&\multicolumn{5}{l}{} 							\\ \hline
\multirow{2}{2.5cm}{Entrapid \cite{Huang.1999.Entrapid}}	&centralized	&kernel			&virtualization	&1999	&No	\\
							&\multicolumn{5}{l}{} 							\\ \hline
\multirow{2}{2.5cm}{Virtual routers \cite{Baumgertner.2002.VirtualRouters}}	&centralized	&kernel		&virtualization	&2003	&No	\\
							&\multicolumn{5}{l}{} 								\\ \hline
\multirow{2}{2.5cm}{GNS3 \cite{gns3}}					&centralized	&kernel			&virtualization	&2008	&\cite{gns3}	\\
							&\multicolumn{5}{l}{-- based on Cisco router images} 			\\ \hline
\multirow{2}{2.5cm}{NETShaper \cite{Herrscher.2002.NetShaper}}		&distributed	&kernel	&event-driven	&2002	&No	\\
							&\multicolumn{5}{l}{-- link-level emulation} 				\\ \hline
\multirow{2}{2.5cm}{Orbit \cite{orbit}}			&distributed	&kernel			&event-driven	&2005	&\cite{orbitweb}	\\
							&\multicolumn{5}{l}{-- two-tier laboratory emulator/field trial network testbed} 	\\ \hline
\multirow{2}{2.5cm}{W-NINE \cite{wnine}}			&distributed	&kernel			&event-driven	&2008	&No	\\
							&\multicolumn{5}{l}{-- wireless emulator} 					\\ \hline
\multirow{2}{2.5cm}{CMU \cite{Noble.1997.Trace-based}}	&centralized	&user			&trace-based	&1997	&No	\\
							&\multicolumn{5}{l}{-- seminal paper on trace-based approach} 			\\ \hline
\end{tabular}
\end{center}
\caption{Summary of the emulators cited in this paper.}
\label{tab}
\end{table}

\section{Conclusion}
\label{Conclusion}

This paper attempts to provide highlights concerning network emulation which is considered to be in the middle between simulation and live-testing schemes.
We saw that network emulation combines the advantages offered by simulation and live-testing at the same time while allowing different evaluation metrics (i.e. processing overhead, memory footprint). Another important finding is that we can easily set up complex measurements testbed by combining both virtualization and network emulation tools.
However network emulation is definitely not the unique answer and must be carefully weighted as a function of the performances an experimenter seeks to evaluate. Thus, to obtain a clear view, we develop an architectural model which illustrates and classifies all types of emulation tools. 
We hope both model and arguments presented would help the reader to better weight emulation in an evaluation process and choose the right scheme to assess the performances targeted.

\bibliographystyle{plain}
\bibliography{Emulation_Survey}

\end{document}